\documentclass[a4paper,10pt]{article}
\usepackage[utf8]{inputenc}
\usepackage{graphicx}
\usepackage{amssymb}
\usepackage{amsmath}

\title{Features of the mass transfer in magnetic cataclysmic variables with fast-rotating white dwarfs}
\author{Polina Isakova$^{1}$\thanks{isakovapb@inasan.ru},
        Andrey Zhilkin$^{1,2}$ and
        Dmitry Bisikalo$^{1}$\\
\textit{\small $^{1}$ Institute of Astronomy RAS, Moscow, Russia}\\%
\textit{\small $^{2}$ Chelyabinsk State University, Chelyabinck, Russia}}
\date{}

\begin{document}

\maketitle

\begin{abstract}
The flow structure in magnetic cataclysmic variables was investigated taking 
into account the effects of strong magnetic field and fast rotation of the white 
dwarf. We modeled the AE Aqr system as a unique object 
that has the rotation period of the white dwarf is about 1000 times shorter than 
the orbital period of the binary system. Observations show that in spite of fast 
rotation of the white dwarf some part of the stream from the inner Lagrange 
point comes into the Roche lobe region. We analyzed possible mechanisms 
preventing material to outflow from the system. 
\end{abstract}

\section{Introduction}
\label{intro}

Magnetic cataclysmic variables (MCVs) are close binary stars, consisting of the donor 
star (red dwarf) and magnetized accreting star (white dwarf) \cite{warner}. 
Generally MCVs can be divided into polars and 
intermediate polars according to the magnitude of their magnetic fields.

Some MCVs have the extremely fast-rotating white 
dwarf. They have the special place in this division. It is called 
''superpropellers'' \cite{zbbUFN}. We can find the intensive outflow of material 
from these systems because of the additional angular momentum from the 
rotating magnetosphere to the material. In this case the accretion disk is not formed. 
The possible representative object of superpropellers is AE Aqr. The analysis 
\cite{Ikhsanov2012} has shown that the most part of the strange behavior of 
this system can be connected with the superpropeller regime of the white dwarf. 

On the other hand, the observational data show that, despite the very fast rotation 
of the magnetosphere in this system, some part of the flow from the inner Lagrange point $L_1$ 
retains in the Roche lobe of the accretor.
In this paper we analyze possible physical mechanisms that can reduce the transfer 
efficiency of the additional angular momentum from the rotating magnetosphere
to the material.

\section{The basic model}
\label{sec-1}

The parameters of the system under consideration were adopted from \cite{Ikhsanov2012} 
and correspond to AE Aqr. The donor star has mass $M_{\rm d} = 0.91~M_{\odot}$ 
and effective temperature $T_{\rm d} = 4000$~K. The accretor star has mass 
$M_{\rm a} = 1.2~M_{\odot}$ and effective temperature $T_{\rm a} = 13000$~K. 
The orbital rotation period is $P_{\rm orb} = 9.88$~h and the orbital separation is 
$A = 3.02~R_{\odot}$. The estimated magnetic field on the white dwarf surface is
$B_{\rm a} = 50$~MG. One more important characteristic is 
the proper rotation period $P_{\rm spin} = 33.08$~s of the white dwarf. 

We use the Cartesian coordinate system ($x$, $y$, $z$) rotating with the orbital 
angular velocity $\Omega = 2\pi / P_{\rm orb}$ around the center of mass of the 
binary system. 
We assume that the accreting star 
has a proper dipolar magnetic field. In the case of asynchronous 
rotation of the accretor its magnetic field ${\bf B}_{*}$ depends on time, but 
it is potential: $\nabla \times {\bf B}_{*} = 0$. 
We can split the total magnetic field 
${\bf B}$ into the background field ${\bf B}_{*}$ of the accretor and the field 
${\bf b}$ induced by currents in the gas dynamic flows, ${\bf B} = {\bf B}_{*} + 
{\bf b}$.

We describe the flow structure using the following equations \cite{zbbUFN}:
\begin{equation}\label{eq-rho1}
 \frac{\partial \rho}{\partial t} + 
 \nabla \cdot \left( \rho {\bf v} \right) = 0,
\end{equation}
\begin{equation}\label{eq-v1}
 \frac{\partial {\bf v}}{\partial t} + 
 \left( {\bf v} \cdot \nabla \right) {\bf v} =  
 -\frac{\nabla P}{\rho} - 
 \frac{{\bf b} \times (\nabla \times {\bf b})}{4 \pi \rho} - 
 \nabla \Phi + 2 \left( {\bf v} \times {\bf \Omega} \right) - 
 {\bf f}_{*},
\end{equation}
\begin{equation}\label{eq-b1}
 \frac{\partial {\bf b}}{\partial t} = 
 \nabla \times \left[ {\bf v} \times {\bf b} +  
 \left( {\bf v} - {\bf v}_{*} \right) \times {\bf B}_{*} - 
 \eta\, \nabla \times {\bf b} \right],
\end{equation}
\begin{equation}\label{eq-s1}
 \rho T \left[ \frac{\partial s}{\partial t} +  
 \left( {\bf v} \cdot \nabla \right) s \right] = 
 n^2 \left( \Gamma - \Lambda \right) + 
 \frac{\eta}{4 \pi} \left( \nabla \times {\bf b} \right)^2,
\end{equation}
where $\rho$ is the density, ${\bf v}$~--- the velocity, $P$~--- the pressure, 
$\Phi$~--- the Roche potential, ${\bf v}_{*}$~--- the velocity of the magnetic 
field lines, $s$~--- the entropy per unit mass, $n = \rho/m_{\rm p}$~--- 
the number density, $m_{\rm p}$~--- the proton mass, $\eta$~--- the magnetic 
diffusivity. The dependencies of the radiative heating $\Gamma$ and cooling 
$\Lambda$ functions on the temperature $T$ are complicated (see \cite{zb2009, 
ZhilkinMM2010} for details). The density, entropy, and pressure are related by 
the equation of state of an ideal gas: $s = c_V \ln (P/\rho^{\gamma})$, where 
$c_V$ is the specific heat capacity of the gas at constant volume and $\gamma = 
5/3$ is the adiabatic index. Note that our model was developed for the 
simulation of plasma flows in strong background magnetic fields. It is based on 
the averaged characteristics of plasma flow in the frame of the wave MHD 
turbulence \cite{zbSMF2010, zbbUFN}. In our approximation the influence of 
the magnetic field of the accretor ${\bf B_*}$ is described by the force 
${\bf f}_*$ (see Eq. \eqref{eq-v1}):
\begin{equation}\label{eq-f1}
 {\bf f}_{*} = \left( {\bf v} - {\bf v}_{*} \right)_{\perp}/t_w,
\end{equation}
where $t_w$ corresponds to the specific time of the penetration of the magnetic 
field into plasma and $\perp$ denotes components of velocity perpendicular to 
${\bf B}_*$. 

\begin{figure*}
\centering
\includegraphics[width=0.45\textwidth]{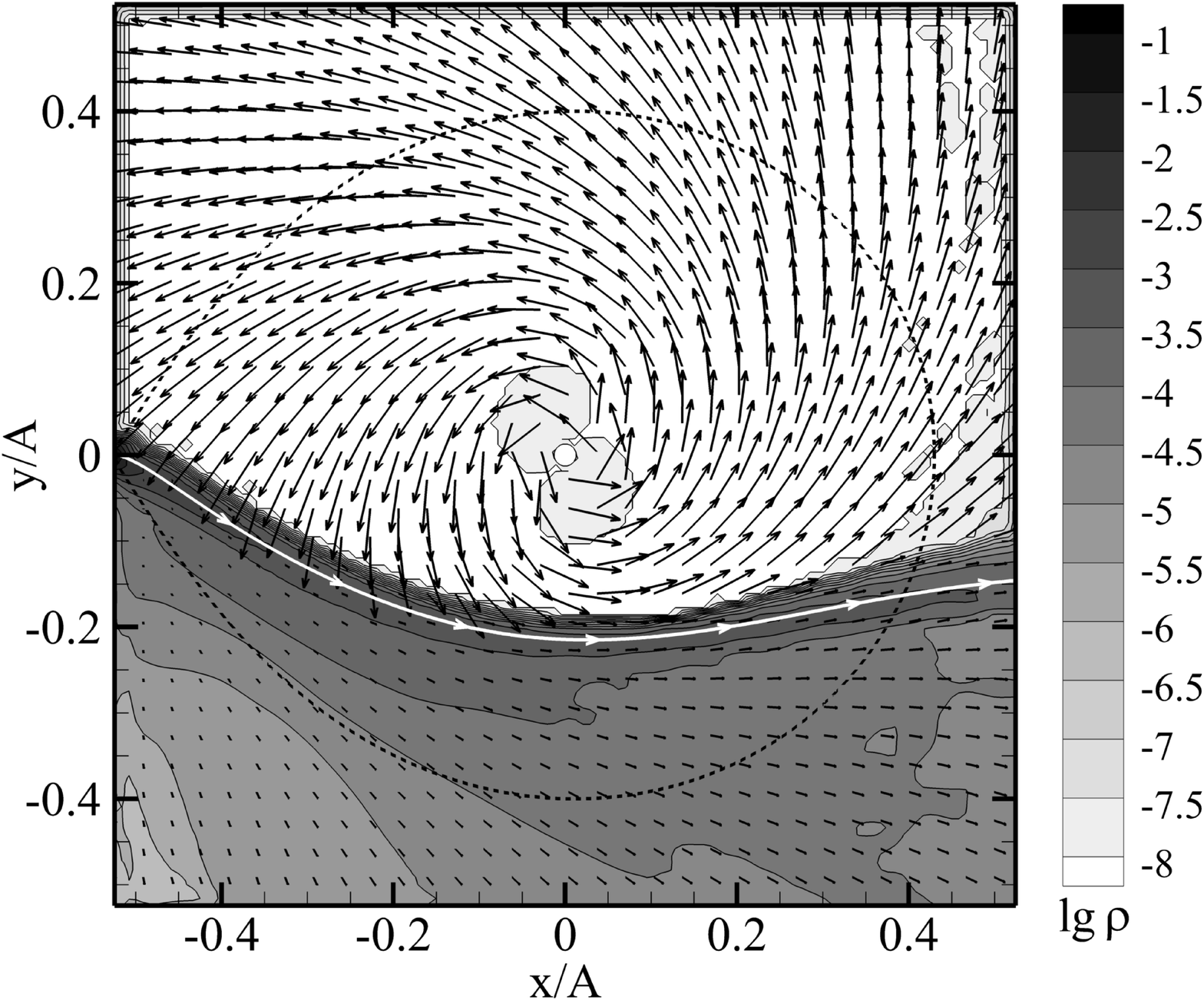}
\includegraphics[width=0.45\textwidth]{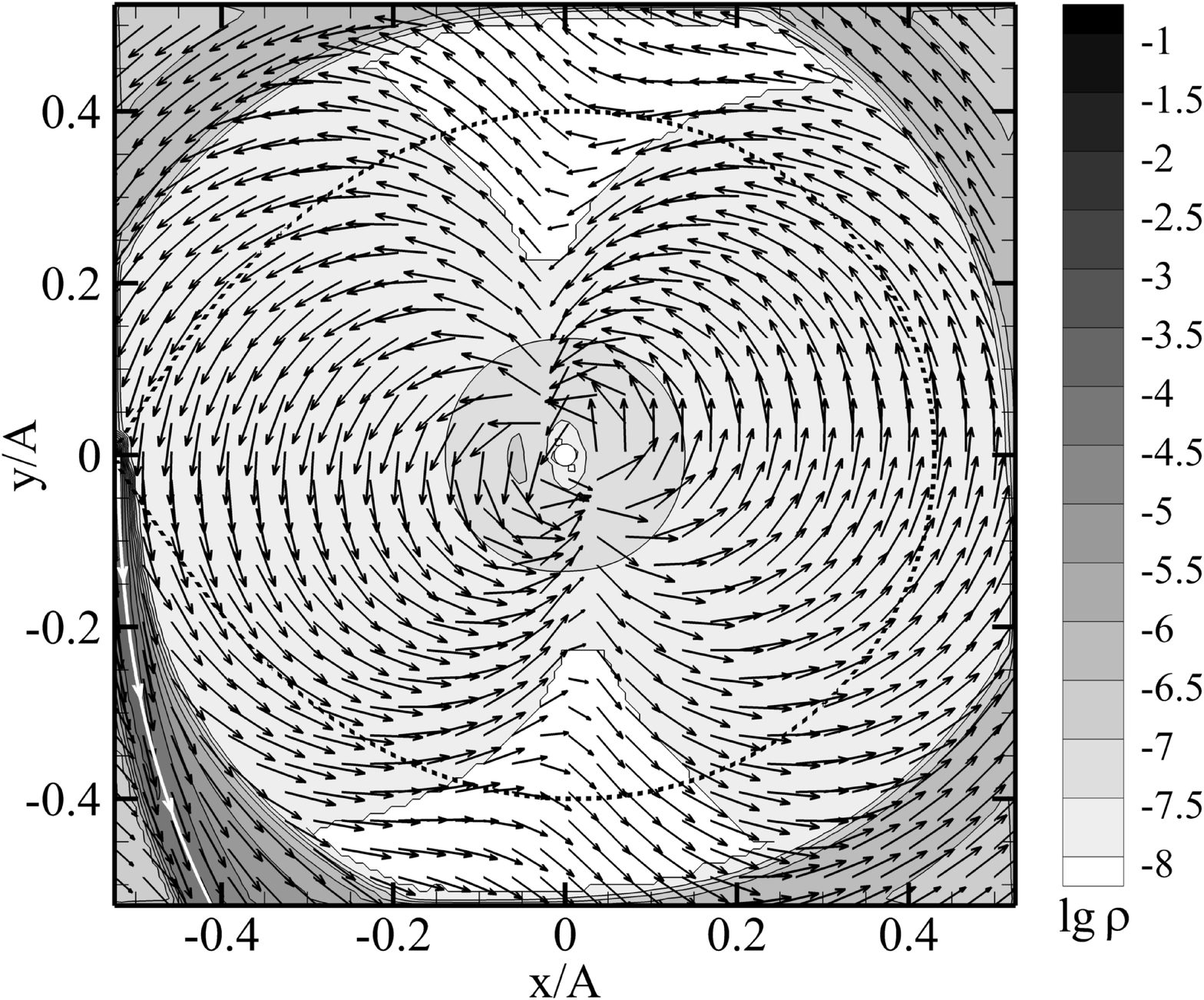}
\caption{The distributions of density (in the logarithmic scale) and velocity vectors
in the equatorial plane of the binary system for the case of the accretor magnetic 
field $0.1$~kG (left panel) and 50~MG (right panel). The dotted line corresponds 
to the Roche lobe boundary. The thick white line with arrows indicates the 
stream line starting from the inner Lagrange point.}
\label{fg1}
\end{figure*}

We use 3D parallel MHD code for the numerical simulations \cite{zb2009, 
ZhilkinMM2010, zbSMF2010}. In the Fig.~\ref{fg1} we present the results of 3D 
simulations in the frame of the basic model for two values of magnetic field 
induction of the accretor. In this approximation the rotation of the white dwarf 
is so fast that there is no accretion in the system. The flowing material is 
just captured by the rotating magnetosphere of the accretor, acquires additional 
angular momentum and flows away from the Roche lobe. The results show that in 
the frame of the basic model the material doesn’t remain in the Roche lobe of 
the accretor even in the case of a relatively weak magnetic field. The similar 
results for the AE Aqr were also received by other authors in the frame of the 
quasi-particle method \cite{Wynn1997, Ikhsanov2004}.

Let us investigate the behavior of the stream in the vicinity of the inner 
Lagrange point $L_1$ in the ballistic approximation. The similar analysis was
done by Lubow and Shu \cite{LubowShu1975} for the non-magnetic case. Therefore
we use their approach but taking into account the magnetic field and rotation 
of the accretor. The formal basis for this assumption is a supersonic flow 
pattern, that allows us to neglect the effects of pressure.

\begin{figure*}
\centering
\includegraphics[width=0.45\textwidth]{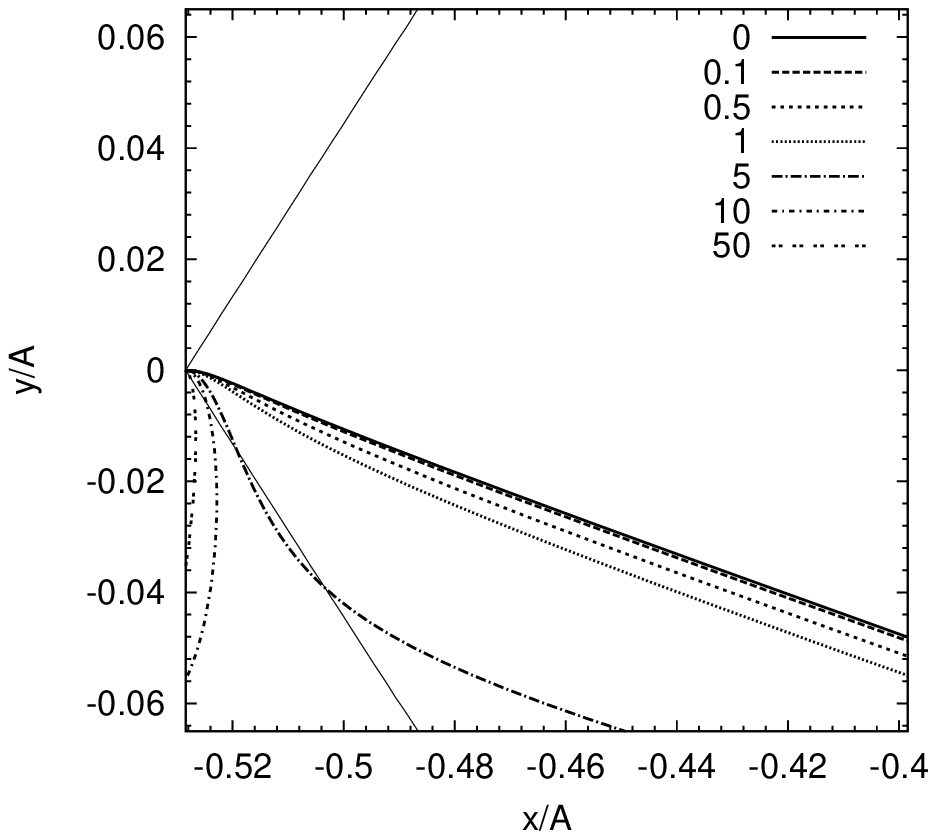}
\includegraphics[width=0.45\textwidth]{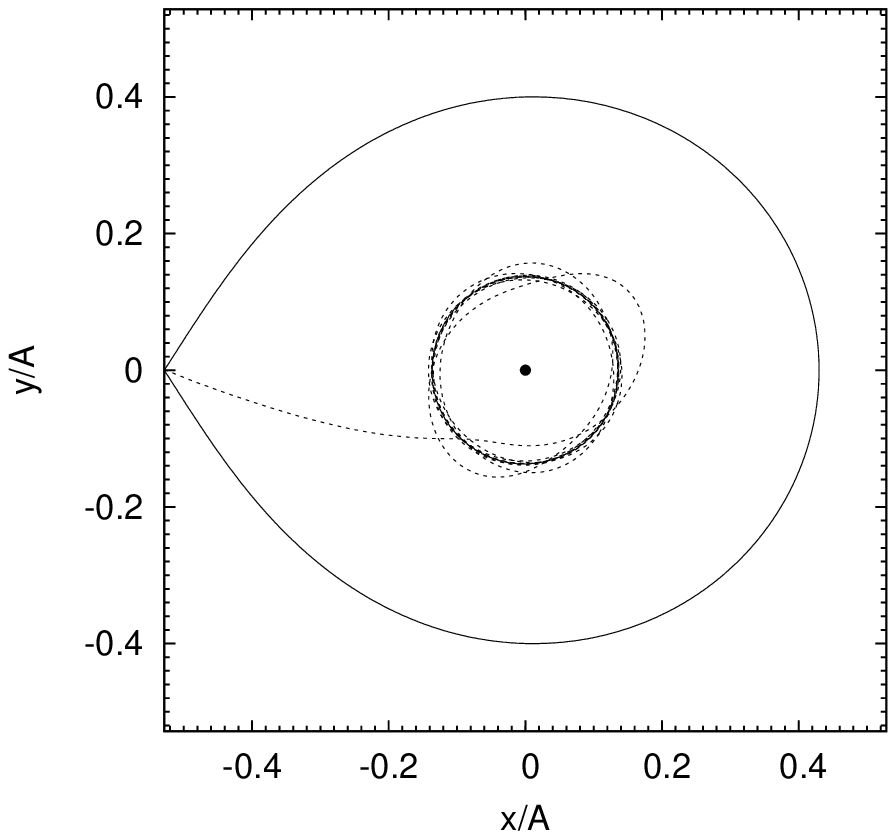}
\caption{The trajectories of test particles in the frame of basic model (left 
panel) and with taking into account relativistic retard of rotating magnetic 
field lines (right panel). On the left panel different curves corresponds to 
different values of the magnetic field induction on the surface of the white 
dwarf $B_{\rm a}$ (in MG). The thin solid line corresponds to the boundary of 
the accretor Roche lobe.}
\label{fg2}
\end{figure*}

For this purpose we can use only the equation of motion \eqref{eq-v1}. In the 
vicinity of the inner Lagrange point $L_1$ this equation can be rewritten in the 
following form:
\begin{equation}\label{eq-v2}
 \frac{d {\bf v}}{d t} = -\nabla\Phi + 
 2 \left( {\bf v} \times {\bf \Omega} \right) - 
 {\bf f}_{*}.
\end{equation}
The left panel of Fig.~\ref{fg2} shows the trajectories of test particles 
obtained in the ballistic approximation in the vicinity of the inner Lagrange 
point $L_1$ for different magnetic fields. The test particles trajectory in pure 
hydrodynamic case is in agreement with the corresponding results of Lubow and 
Shu \cite{LubowShu1975}. The other trajectories for the case of the non-zero 
magnetic field confirm the results of our 3D MHD simulations. For the weak 
magnetic field ($B_{\rm a} < 1$~MG) the trajectories are not strongly different 
from the non-magnetic case. But in the case of strong magnetic fields 
($B_{\rm a} \ge 1$~MG) the trajectories significantly deviate from the pure 
hydrodynamic trajectory and for $B_{\rm a} \ge 5$~MG even do not come to the 
Roche lobe of the white dwarf. Our calculations show that for the widely adopted 
parameters of the magnetic field in the AE Aqr system the material almost 
immediately flows away from the Roche lobe of the white dwarf. So we need to 
consider some physical mechanisms keeping material in the Roche lobe.

\section{Possible mechanisms of the material retention}
\label{sec-2}

\subsection{The partial ionization of the plasma}
\label{sec-21}

One of the mechanisms can be connected with the partial ionization of plasma. 
In fact, in the case of very week degree of ionization we should obtain the 
hydrodynamic solution when an accretion disk forms in the system. In the case of 
partial ionization of plasma the effect of ambipolar diffusion can be very 
effective. The correspondent solution should be intermediate between pure 
hydrodynamic and ideal MHD ones. 
After some simple 
manipulations we can obtain the following expression for averaged 
electromagnetic force acting on plasma from the external magnetic field:
\begin{equation}\label{eq-f2}
 {\bf f}_{*} =  
 -\frac{1}{1 + \chi^2} 
 \frac{({\bf v} - {\bf v}_{*})_{\perp}}{t_*} - 
 \frac{\chi}{1 + \chi^2} 
 \frac{{\bf n}_{*} \times ({\bf v} - {\bf v}_{*})}{t_*}, 
\end{equation}
where ${\bf n}_*$ is the unit vector along the external magnetic field 
${\bf B}_*$, 
\begin{equation}\label{eq-ad1}
 t_* = t_w + \frac{1 - x}{x}\, \tau_i, \quad
 \chi = \frac{1}{x \omega_p t_*},
\end{equation}
$\omega_{\rm p} = e B_* / (m_{\rm p} c)$ describes Larmor frequency for protons, 
$x$ is the degree of ionization, $\tau_i$~--- characteristic time between 
collisions of ions with neutrals. 

Analysis of the corresponding solutions shows that the effect of ambipolar 
diffusion can play an important role in the system for very low degrees of 
ionization in a range from $10^{-12}$ to $10^{-14}$. But the estimation of the 
equilibrium degree of ionization from the Saha equation gives values in a range 
from $10^{-2}$ to $10^{-3}$ in the inner Lagrange point $L_1$ of the system AE Aqr. 
So this mechanism doesn't work in the considered system.

\subsection{Pressure of the magneto-dipole radiation}
\label{sec-22}

The other mechanism can be connected with relativistic effects in the system 
due to the very fast rotation of the accretor. To estimate these 
possible effects we should consider the electromagnetic field of the rotating 
magnetic dipole in the vacuum. The corresponding solution of Maxwell equations 
in the laboratory rest frame can be written in the form:
\begin{equation}\label{eq-m1}
 {\bf E} = 
 \frac{1}{r^2 c} 
 \left( {\bf n} \times \dot{{\bf m}} \right) + 
 \frac{1}{r c^2} \left( {\bf n} \times \ddot{{\bf m}} \right),
\end{equation}  
\begin{equation}\label{eq-m2}
 {\bf B} = 
 \frac{1}{r^3} 
 \left[ 3 {\bf n} ( {\bf n} \cdot {\bf m} ) - {\bf m} \right] + 
 \frac{1}{r^2 c} 
 \left[ 3 {\bf n} \left( {\bf n} \cdot \dot{{\bf m}} \right) - 
 \dot{{\bf m}} \right] + 
 \frac{1}{r c^2} 
 \left[ {\bf n} \times \left( {\bf n} \times \ddot{{\bf m}} \right) \right],
\end{equation}
where ${\bf n} = {\bf r}/r$ and ${\bf m} = {\bf m}(t - r/c)$ is the vector of 
magnetic moment. 

The characteristic spatial scale for the electromagnetic field \eqref{eq-m1}, 
\eqref{eq-m2} is the light radius $r_l = c P_{\rm spin}/(2\pi)$. In the AE Aqr 
system the boundary of the light cylinder is located farther than the inner 
Lagrange point but closer than the center of the donor. So the relativistic 
effects actually can play an important role in this system. 

We can consider the pressure of the magneto-dipole radiation as one of the 
mechanisms that can keep the material in the Roche lobe of the accretor. 
To include this effect we should add the corresponding force 
determined by the last terms in Eqs.  \eqref{eq-m1} and \eqref{eq-m2} into our 
numerical model. But simple estimations show that in our case in the Lagrange point 
$L_1$ this force is about $10^2$--$10^3$ times smaller than the gravitational 
force of the accretor. Therefore we can neglect this effect, too.

\subsection{The relativistic retard}
\label{sec-23}

Another relativistic effect is the retard of rotating magnetic field lines. In 
fact, in relativistic case the electromagnetic field of the rotating magnetic 
dipole \eqref{eq-m1}, \eqref{eq-m2} is determined by the magnetic moment 
${\bf m}$ at a retarded time $t' = t - r/c$. In particular, this means that the 
magnetic field lines rotate not like a solid body. The retard effect increases 
with the distance from the acrretor. At the boundary of the light cylinder the 
velocity of the field lines rotation equals the speed of light $c$. Therefore 
outside the light cylinder the magnetic field in exact solution cannot transfer 
effectively the angular momentum to the material. In our model we can describe 
roughly this effect using an empirical dependence for angular velocity of 
magnetic lines on radius. 

\begin{figure*}
\centering
\includegraphics[width=0.45\textwidth]{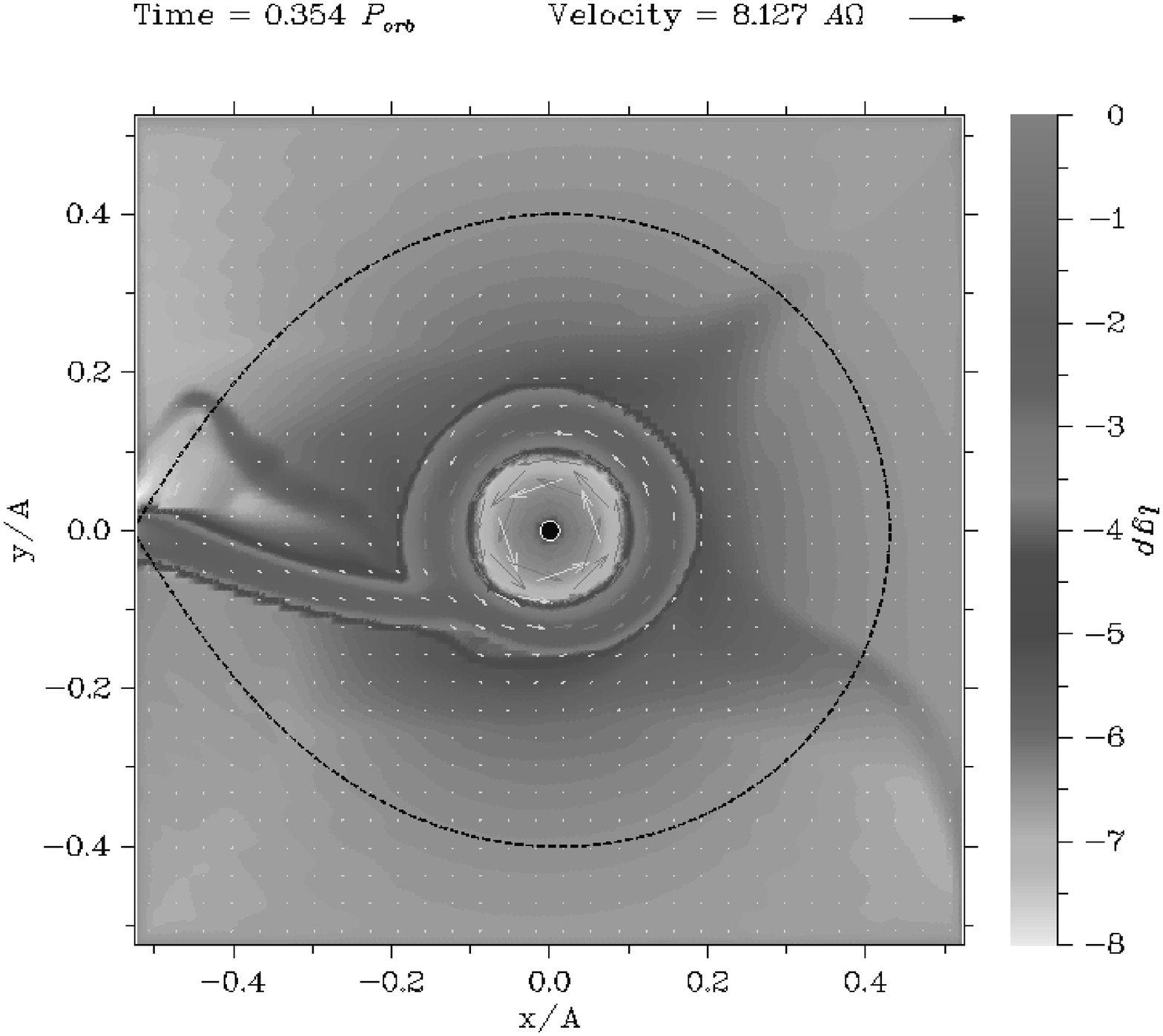}
\includegraphics[width=0.45\textwidth]{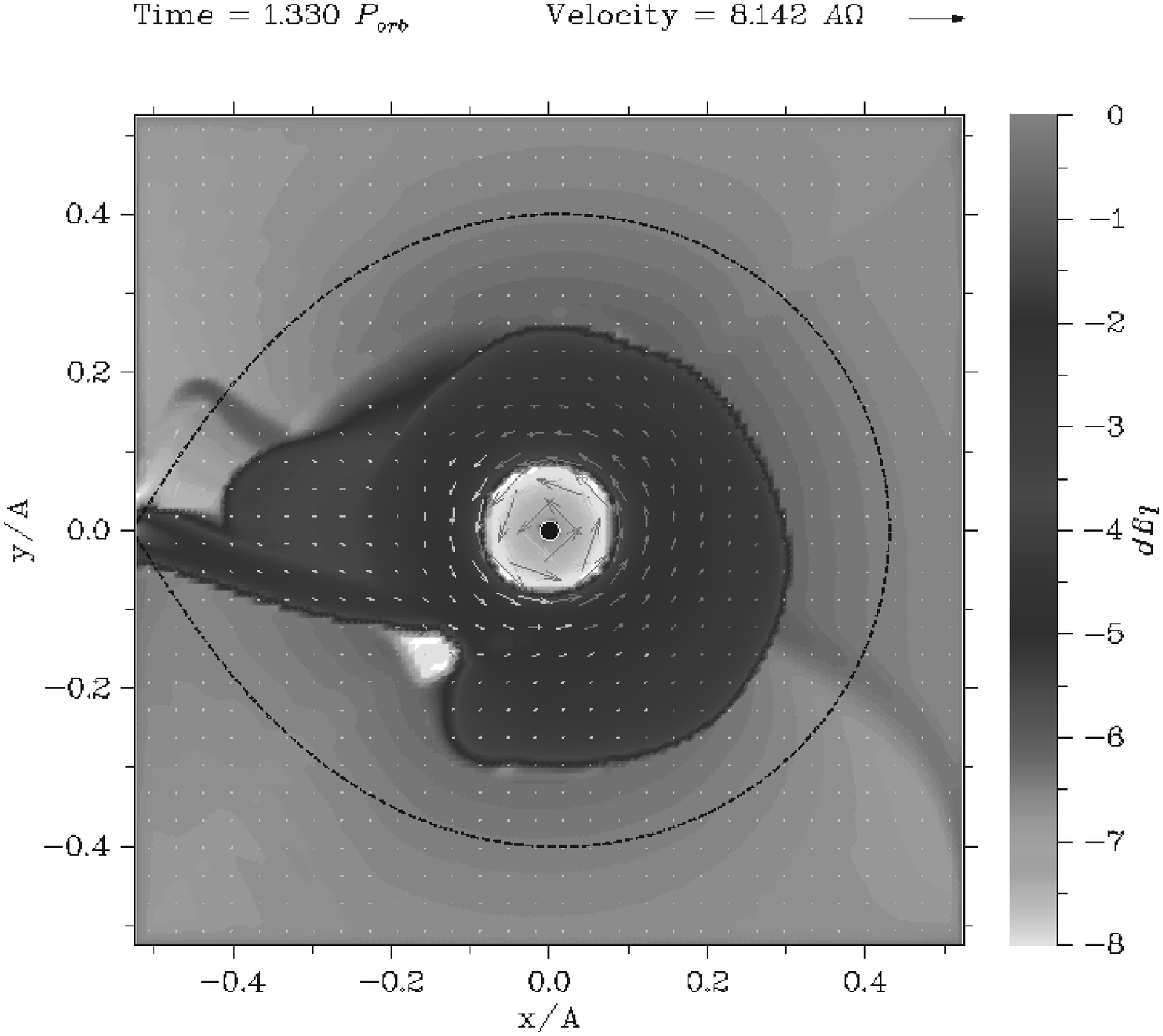}
\caption{The distributions of density and velocity vectors for different time 
moments in the equatorial ($xy$) plane of the binary system. The dashed line 
corresponds to the Roche lobe boundary.}
\label{fg3}
\end{figure*}

The right panel of the Fig. \ref{fg2} shows a result of calculations in the 
ballistic approximation in the case of exponential dependence for angular 
velocity of magnetic lines rotation on radius. It is seen that in the frame of this 
approach we can easy find the model parameters when the sufficient part of 
material stays in the Roche lobe of the white dwarf. 

The results of 3D simulations are shown in Fig. \ref{fg3}. The figure 
demonstrates the distribution of the density 
and the velocity field in the equatorial plane of the binary system.
We can see that at initial stage (left panel of Fig. \ref{fg3}) the inflowing 
material forms a ring around the accretor. The gradual accumulation of material 
in the ring leads to increase of the pressure gradient. Therefore the subsequent 
evolution leads to the transformation of the original ring into a disk. At a 
particular time moment (right panel of Fig. \ref{fg3}) the material overfills 
the Roche lobe and quasi-periodical overflows begin. So, the relativistic retard 
of rotating magnetic field lines can be considered as an effective mechanism 
for retention of material in the Roche lobe of the accretor in MCVs with fast 
rotating magnetized white dwarfs.

\section{Conclusions}
\label{sec-3}

In this work we investigate the flow structure in MCVs taking into account
the effects of strong magnetic field and fast 
rotation of the white dwarf. The AE Aqr system is considered as an example. 
We analyzed possible physical mechanisms that can prevent 
the material to outflow from the binary system. Among these mechanisms are the 
ambipolar diffusion, the pressure of magneto-dipole radiation and the 
relativistic retard of rotating magnetic field lines. 

Our analysis has shown that the last mechanism can be the most effective. Both 
ballistic calculations and MHD 3D simulations show the formation of fairly 
stable ring-like structures around the accretor in the AE Aqr system caused by 
presence of the relativistic retard of rotating magnetic field lines. The 
gradual accumulation of material in the ring leads to increase of the pressure 
gradient, and, as a result, to the transforming of the original ring into the 
disk.\\

The work was supported by the Basic Research Program of the Presidium of the 
Russian Academy of Sciences, Russian Foundation for Basic Research (projects 
11-02-00076, 12-02-00047, 13-02-00077, 13-02-00939), Federal Targeted Program 
''Science and Science Education for Innovation in Russia 2009-2013''.

\end{document}